\documentclass[aps,pra,twocolumn,showpacs,superscriptaddress,nofootinbib]{revtex4-1}
\usepackage{amsmath}
\usepackage{amssymb}
\usepackage{graphicx}
\usepackage{mathrsfs}
\usepackage{ntheorem}
\usepackage{enumerate}
\usepackage{enumitem}
\usepackage{times,txfonts}

\newcommand{\ket}[1]{|#1\rangle}
\newcommand{\bra}[1]{\langle #1|}
\newcommand{\Tr}{\mathrm{Tr}}

\newtheorem{theorem}{Theorem}
\newtheorem{lemma}{Lemma}

\def\CC{{\rm\kern.24em \vrule width.04em height1.46ex depth-.07ex \kern-.30em C}}
\def\RR{{\rm\kern.24em \vrule width.04em height1.46ex depth-.07ex
\kern-.30em R}}
\def\P{{\rm I\kern-.25em P}}

\begin{document}
\title{General approach to find steady-state manifolds in Markovian and non-Markovian systems}
\author{Da-Jian Zhang}
\affiliation{School of Mathematics, Shandong University, Jinan 250100, China}
\affiliation{Department of Physics, Shandong University, Jinan 250100, China}
\author{Xiao-Dong Yu}
\affiliation{Department of Physics, Shandong University, Jinan 250100, China}
\author{Hua-Lin Huang}
\affiliation{School of Mathematical Sciences, Huaqiao University, Quanzhou 362021, China}
\author{D. M. Tong}
\email{tdm@sdu.edu.cn}
\affiliation{Department of Physics, Shandong University, Jinan 250100, China}
\date{\today}
\begin{abstract}
Steady-state manifolds of open quantum systems, such as decoherence-free subspaces and noiseless subsystems, are of great practical importance to the end of quantum information processing. Yet, it is a difficult problem to find steady-state manifolds of open quantum systems, especially of non-Markovian systems. In this paper, we propose an approach to find the steady-state manifolds, which is generally applicable to both Markovian and non-Markovian systems. Our approach is based on an arbitrarily given steady state, and by following the standard steps of the approach, the steady-state manifold on the support subspace of the given state can be obtained. Our work reduces the problem of finding a manifold of steady states to that of finding only one steady state, which is indeed an interesting progress towards completely solving the difficult problem. Besides, in deriving our approach, we introduce the notions of the modified noise algebra and its commutant, and prove two theorems on the structure of steady-state manifolds of general open systems, which themselves are interesting findings too.
\end{abstract}
\maketitle

\section{Introduction}

Any real quantum system inevitably interacts with its environment, and such an interaction generally spoils coherence of quantum states and further weakens the abilities of quantum states to perform quantum information processing tasks. A challenge in quantum information processing is to overcome decoherence of quantum states caused by undesired interactions. Interestingly, despite such destructive interactions, there are steady states in many open systems, which remain stable during the evolutions of the systems and are completely immune to decoherence. Due to their coherence stabilization virtues, steady states of open systems have been exploited for various aims of quantum information processing, such as quantum error correction
\cite{Zanardi1997,Lidar1998,Ollerenshaw2003,Mohseni2003,Viola2000,Zanardi2000,Viola2001},
quantum state preparation \cite{Diehl2008,Kastoryano2011,Krauter2011,Torre2013,Carr2013,Znidari2016}, quantum computation \cite{Wu2005,Oreshkov2009,Xu2012,Feng2013,Xu2014,ZhangJ2014,Verstraete2009,Zanardi2014},  quantum simulation \cite{Barreiro2011,Stannigel2014}, entanglement distillation \cite{Vollbrecht2011}, and cooling \cite{Mari2012}.

The practical importance of steady states for realizing quantum information processing leads to a surge of interest in finding steady states of open systems. Recently, authors have put forward several methods for finding steady states of open systems, such as the mean-field method \cite{Diehl2010,Lee2013,Jin2013}, the variational method based on the minimization of trace norm \cite{Weimer2015}, and the variational method based on the matrix product operator ansatz \cite{Cui2015}. These methods have been successfully applied to many well-known physical models, e.g., the Dicke model, the dissipative Ising model, the dissipative Bose-Hubbard model, and the strongly interacting Rydberg gases.

However, the problem is that all these previous methods are only applicable to Markovian systems but not applicable to non-Markovian systems, and there has not been a general approach to find the steady states of non-Markovian systems. Besides, the previous methods mainly focus on finding individual steady states, but cannot give the structure of steady-state manifolds (SSMs). Yet, for open systems with many or infinitely many steady states, the knowledge about the structure of SSMs, such as decoherence-free subspaces \cite{Zanardi1997,Lidar1998} and noiseless subsystems \cite{Viola2000,Zanardi2000}, plays a key role in performing many quantum information processing tasks \cite{Zanardi2014,Zanardi2015,Zhang2016}. Therefore, it is an important issue to find SSMs of general open systems and determine the structure of them.

In the present paper, we address this issue. We first introduce the modified noise algebra and its commutant, with which we can obtain two theorems on the structure of SSMs, and based on the theorems, we further propose an approach to find the SSMs. Our approach, based on a given steady state, is applicable to both Markovian and non-Markovian systems. By following our approach, the SSM on the support subspace of the given steady state can be found. This paper is organized as follows. In Sec. \ref{sec2}, we introduce the modified noise algebra and its commutant, and present a theorem on the relation between them and the SSM. In Sec. \ref{sec3}, we give a theorem on the structure of the SSMs with the help of the modified noise algebra. In Sec. \ref{sec4}, based on the the structure theorem, we put forward an approach for finding SSMs.  Section \ref{sec5} provides three examples to illustrate the usefulness of our approach, and Sect. \ref{sec6} presents the summary and remarks.

\section{Modified noise algebra, commutant algebra, and the relation between them and steady-state manifold}\label{sec2}

We first specify some notations and terminologies. $\mathcal{E}_t$ is used to denote the dynamics of an open system, i.e., a completely positive and trace-preserving (CPTP) map, which transforms the initial state $\rho$ to the state at time $t$, $\rho(t)=\mathcal{E}_t(\rho)$. A CPTP map can be always written as the Kraus representation, $\mathcal{E}_t(\rho)=\sum_kE_k(t)\rho E_k^\dagger(t)$, with Kraus operators $E_k(t)$ satisfying $\sum_kE_k^\dagger(t)E_k(t)=I$. $\mathcal{H}$ ($\textrm{dim}\mathcal{H}<\infty$) is used to represent the Hilbert space for the system and $L(\mathcal{H})$ the algebra of linear operators on it.

A steady state is such a state which satisfies $\mathcal{E}_t(\rho)=\rho$ for all the time $t$, and a strict SSM is a set of steady states. A generalized definition of an SSM comprises operators that satisfy $\mathcal{E}_t(X)=X$ for all the time $t$. Here, in our paper, we adopt the generalized definition of SSMs, $\{X\in L(\mathcal{H})|\mathcal{E}_t(X)=X,\forall t\}$, in which the elements $X$ are not necessarily density operators. Such a generalized definition is convenient for serving our purpose.

We introduce the notions of noise algebra and its commutant. The noise algebra, denoted as $\mathcal{A}$, is the algebra generated by all Kraus operators and their conjugate transposes, of which the elements are linear combinations of products of the operators in $\{E_k(t), E_{k}^\dagger(t), \forall k,t\}$. The commutant of a noise algebra, denoted as $\mathcal{A}^\prime$,  is the set of the elements that commute with each element of the noise algebra. It is worth noting that the noise algebra as well as its commutant is only dependent on the CPTP map $\mathcal{E}_t$ but independent of the choices of Kraus operators of the map.

Our approach is based on a given steady state, i.e., it is supposed that one steady state, denoted as $\rho_0$,  has been known. The support of $\rho_0$ is denoted as $P_{\rho_0}$, which is the smallest orthogonal projection operator that satisfies $\Tr(\rho_0 P_{\rho_0})=1$. $P_{\rho_0}\mathcal{H}$ represents the subspace defined by the operator $P_{\rho_0}$, and there is  $\mathcal{H}=P_{\rho_0}\mathcal{H}\oplus(I-P_{\rho_0})\mathcal{H}$. Our task is to find all the steady states (i.e., the SSM) supported on $P_{\rho_0}\mathcal{H}$, $\{X\in L(P_{\rho_0}\mathcal{H})|\mathcal{E}_t(X)=X,\forall t\}$.

With the above notations and terminologies, we start our discussion on the approach of finding SSMs.

To prepare for general open systems, we first consider an important family of open systems that are with unital CPTP maps, $\mathcal{E}_t(I)=I$. In this case,  $\rho_0=cI$, where $c$ is the normalization constant, is always a steady state. For these open systems, we have the following lemma.

\begin{lemma}\label{lemma1}
The steady-state manifold of the open system under a unital CPTP map is equal to the commutant of the noise algebra.
\end{lemma}

Lemma \ref{lemma1} is easy to be proved. Since $\mathcal{E}_t$ is unital, i.e.,  $\mathcal{E}_t(I)=\sum_kE_k(t)E_k^\dagger(t)=I$, there is $\mathcal{E}_t(X)=\sum_kE_k(t)XE_k^\dagger(t)=X\sum_kE_k(t)E_k^\dagger(t)=X$ for all the operators $X\in\mathcal{A}^\prime$, which means $\textrm{SSM}\supset\mathcal{A}^\prime$. Hence, we only need to prove $\textrm{SSM}\subset\mathcal{A}^\prime$. To this end, we let $X$ belong to SSM. Then, $X^\dagger$ belongs to SSM, too, since $\mathcal{E}_t$ is completely positive and hence Hermitian-preserving, i.e., $\mathcal{E}_t(X^\dagger)=\mathcal{E}_t(X)^\dagger$. By using the Schwarz type inequality \cite{Lieb1974}, we have $\mathcal{E}_t(XX^\dagger)\geq \mathcal{E}_t(X)\mathcal{E}_t(X^\dagger)$, which further leads to $\mathcal{E}_t(XX^\dagger)\geq XX^\dagger$, i.e., $\mathcal{E}_t(XX^\dagger)-XX^\dagger$ is positive semidefinite. Since  $\mathcal{E}_t$ is a trace-preserving map, we have  $\Tr[\mathcal{E}_t(XX^\dagger)-XX^\dagger]=0$. This expression is valid if and only if $\mathcal{E}_t(XX^\dagger)-XX^\dagger=0$, i.e., $XX^\dagger\in\textrm{SSM}$. Using $\mathcal{E}_t(I)=I$ and $X,X^\dagger,XX^\dagger\in\textrm{SSM}$, we have
$X\mathcal{E}_t(I)X^\dagger-X\mathcal{E}_t(X^\dagger)-\mathcal{E}_t(X)X^\dagger+\mathcal{E}_t(XX^\dagger)=0$.
Rewriting this equation gives $\sum_k[X,E_k(t)][X,E_k(t)]^\dagger=0$. Since $\sum_k[X,E_k(t)][X,E_k(t)]^\dagger$ is a sum of nonnegative terms, each of the terms in it must be zero, which implies $[X,E_k(t)]=0$ for all $k$ and $t$. Replacing $X$ with $X^\dagger$ in the foregoing arguments, we have $[X^\dagger,E_k(t)]=0$, and hence $[X,E_k^\dagger(t)]=0$, for all $k$ and $t$, too. It follows that $X\in\mathcal{A}^\prime$. This completes the proof of Lemma \ref{lemma1}.

This lemma reveals the relation between the SSM and the noise algebra for open systems under unital CPTP maps. It shows that a nontrivial SSM of an open  system with a unital CPTP map exists if and only if the commutant of the noise algebra is nontrivial. With the result about unital CPTP maps, we may now turn to investigate open systems under general CPTP maps. We will give a generalized relation between SSM and the noise algebra for open systems under general CPTP maps.

We now consider open systems under general CPTP maps $\mathcal{E}_t$. Our discussion is based on the assumption that a steady state $\rho_0$ has been known and we aim to find the SSM supported on the subspace $P_{\rho_0}\mathcal{H}$. Since $P_{\rho_0}\mathcal{H}$ is invariant under the action of $\mathcal{E}_t$ \cite{Note}, it is enough to consider the subspace $P_{\rho_0}\mathcal{H}$. From now on, we will assume that the operators appearing in the following are redefined to act on $P_{\rho_0}\mathcal{H}$ instead of $\mathcal{H}$. We define a modified noise algebra and its commutant, denoted as $\mathcal{A}_{\rho_0}$ and $\mathcal{A}_{\rho_0}^\prime$, respectively. $\mathcal{A}_{\rho_0}$  is generated by the modified Kraus operators $\tilde{E}_k(t):=\rho_0^{-\frac{1}{2}}E_k(t)\rho_0^{\frac{1}{2}}$ and $\tilde{E}_k^\dagger(t)$, instead of $E_k(t)$ and $E_k^\dagger(t)$ in the noise algebra $\mathcal{A}$.  That is, the elements of $\mathcal{A}_{\rho_0}$ are linear combinations of products of the operators in $\{\tilde{E}_k(t), \tilde{E}_{k}^\dagger(t), \forall k,t\}$. $\mathcal{A}_{\rho_0}^\prime$  is the set of the elements that commute with each element of the modified noise algebra $\mathcal{A}_{\rho_0}$. Then, we have the following theorem.

\begin{theorem}\label{th1}
The steady-state manifold supported on the subspace $P_{\rho_0}\mathcal{H}$ is equal to $\rho_0^{\frac{1}{2}}\mathcal{A}_{\rho_0}^\prime\rho_0^{\frac{1}{2}}$.
\end{theorem}

Theorem \ref{th1} can be proved with the aid of Lemma \ref{lemma1}. To this end, we define an ancillary completely positive map, denoted as $\tilde{\mathcal{E}}_t$, by the Kraus operators $\tilde{E}_k(t)$, namely, $\tilde{\mathcal{E}}_t(X)=\sum_k\tilde{E}_k(t)X\tilde{E}_k^\dagger(t)$.  The dual map of $\tilde{\mathcal{E}}_t$ is denoted as $\tilde{\mathcal{E}}_t^*$, which is defined by the relation $\langle \tilde{\mathcal{E}}_t^*(X), Y\rangle=\langle X, \tilde{\mathcal{E}}_t(Y)\rangle$ for $X$, $Y$ $\in L(P_{\rho_0}\mathcal{H})$, where $\langle X, Y\rangle=\Tr({X^\dagger Y})$ is the Hilbert-Schmidt inner product \cite{Nielsen2000}. It is easy to verify that the ancillary map $\tilde{\mathcal{E}}_t$  has the following properties:
(P1) $\tilde{\mathcal{E}}_t$ is unital, i.e., $\tilde{\mathcal{E}}_t(P_{\rho_0})=P_{\rho_0}$;
(P2) $\rho_0$ is a steady state of $\tilde{\mathcal{E}}_t^*$, i.e., $\tilde{\mathcal{E}}_t^*(\rho_0)=\rho_0$;
(P3) $\tilde{\mathcal{E}}_t(\tilde{\rho})=\tilde{\rho}$ if and only if $\mathcal{E}_t(\rho)=\rho$, where $\rho=\rho_0^{\frac{1}{2}}\tilde{\rho}\rho_0^{\frac{1}{2}}$.

Property (P3) implies that the steady states supported on the subspace $P_{\rho_0}\mathcal{H}$ under $\mathcal{E}_t$ and those under  $\mathcal{\tilde{E}}_t$  are corresponding to each other, namely, $\{X\in L(P_{\rho_0}\mathcal{H})|\mathcal{E}_t(X)=X,\forall t\}= \{\rho_0^{\frac{1}{2}}X\rho_0^{\frac{1}{2}}|\mathcal{\tilde{E}}_t(X)=X,\forall t\}$.

With the aid of the ancillary map $\mathcal{\tilde{E}}_t$ and Lemma \ref{lemma1}, it is easy to prove Theorem \ref{th1}. In fact, by following the same arguments as those in the proof of Lemma \ref{lemma1} and using the properties (P1) and (P2), we can prove that the set $\{X\in L(P_{\rho_0}\mathcal{H})|\tilde{\mathcal{E}}_t(X)=X,\forall t\}$ is equal to the commutant of the modified noise algebra $\mathcal{A}_{\rho_0}^\prime$.
To do this, we only need to replace $\mathcal{E}_t$, $E_k(t)$, and the equation $\Tr[\mathcal{E}_t(XX^\dagger)-XX^\dagger]=0$ in the proof of Lemma \ref{lemma1} with $\mathcal{\tilde{E}}_t$, $\tilde{E}_k(t)$, and the equation $\Tr\{\rho_0[\mathcal{\tilde{E}}_t(XX^\dagger)-XX^\dagger]\}=0$, respectively. Then, we have from (P3) that the SSM supported on the subspace $P_{\rho_0}\mathcal{H}$ is equal to $\rho_0^{\frac{1}{2}}\mathcal{A}_{\rho_0}^\prime\rho_0^{\frac{1}{2}}$. This completes the proof of Theorem \ref{th1}.

Based on a given steady state $\rho_0$, we have established the relation between the SSM supported on $P_{\rho_0}\mathcal{H}$ and the modified noise algebra. Theorem \ref{th1} indicates that a nontrivial SSM exists if and only if the commutant of the modified noise algebra is nontrivial. It presents a unified picture of the relation between SSMs and noise algebras, which is applicable to unital CPTP maps as well as non-unital CPTP maps, and  Markovian systems as well as non-Markovian systems. Particularly, for unital CPTP maps, $\rho_0$ can be taken as $cI$, and there are $\tilde{E}_k(t)=E_k(t)$, $\mathcal{\tilde{E}}_t=\mathcal{E}_t$, and $\mathcal{A}_{\rho_0}=\mathcal{A}$. Consequently, Theorem \ref{th1} reduces to Lemma \ref{lemma1}. In the case of a Markovian system, the dynamics is described by the Lindblad equation, $\partial_t\rho(t)=\mathcal{L}\rho=-i[H,\rho]+\sum_k A_k\rho A_k^\dagger-\frac{1}{2}\{A_k^\dagger A_k,\rho\}$, with the time-independent Hamiltonian $H$ and time-independent Lindblad operators $A_k$~\cite{Lindblad1976}. The Lindblad equation is equivalent to the Kraus representation with the Kraus operators $E_0(dt):=I-\left(iH+\frac{1}{2}\sum_kA_k^\dagger A_k\right)dt$ and $E_k(dt):=A_k\sqrt{dt}$. Indeed, direct calculations show that $\rho(t+dt):=E_0(dt)\rho(t)E_0^\dagger(dt)+\sum_kE_k(dt)\rho(t)E_k^\dagger(dt)
=\rho(t)+\mathcal{L}\rho(t)dt+O({dt}^2)$, from which the equivalence between the Lindblad equation and this Kraus representation can be proved. Using the equivalent Kraus representation, we can obtain, from Theorem \ref{lemma1}, that the SSM supported on the subspace $P_{\rho_0}\mathcal{H}$ is equal to $\rho_0^{\frac{1}{2}}\textrm{alg}\{\tilde{H},\tilde{H}^\dagger,\tilde{A}_k,\tilde{A}_k^\dagger,\forall k\}^\prime\rho_0^{\frac{1}{2}}$, where $\tilde{H}=\rho_0^{-\frac{1}{2}}H\rho_0^{\frac{1}{2}}$, $\tilde{A}_k=\rho_0^{-\frac{1}{2}}A_k\rho_0^{\frac{1}{2}}$, and $\textrm{alg}\{\tilde{H},\tilde{H}^\dagger,\tilde{A}_k,\tilde{A}_k^\dagger,\forall k\}$ represents the algebra generated by $\tilde{H}$, $\tilde{A}_k$, and their conjugate transposes.

\section{Structure of steady-state manifolds}\label{sec3}

We now work out the structure of the SSM supported on the subspace $P_{\rho_0}\mathcal{H}$. To this end, we need first to prove the following lemma.

\begin{lemma}\label{lemma2}
A steady state with the same support as $\rho_0$ always exists in the modified noise algebra $\mathcal{A}_{\rho_0}$.
\end{lemma}

Lemma \ref{lemma2} means that among all the steady states that have the same support as $\rho_0$, there is one of them belonging to the modified noise algebra $\mathcal{A}_{\rho_0}$. To prove this, we let $\overline{\rho}_0:=\int U\rho_0U^\dagger dU$, where the Haar integral runs over all the unitary operators in $\mathcal{A}_{\rho_0}^\prime$. First, it is obvious that $\overline{\rho}_0$  has the same support as $\rho_0$. Second, $\overline{\rho}_0$ is a steady state. This is because $\mathcal{E}_t(U\rho U^\dagger)=U\mathcal{E}_t(\rho)U^\dagger$ for all $\rho\in L(P_{\rho_0}\mathcal{H})$ and $U\in\mathcal{A}_{\rho_0}^\prime$ (See Appendix \ref{appA}), which leads to $\mathcal{E}_t(\overline{\rho}_0)=\int \mathcal{E}_t(U\rho_0U^\dagger) dU=\int U\mathcal{E}_t(\rho_0)U^\dagger dU=\int U\rho_0U^\dagger dU=\overline{\rho}_0$. Third, we can show that $\overline{\rho}_0\in\mathcal{A}_{\rho_0}$. To do this, letting $U^\prime\in\mathcal{A}_{\rho_0}^\prime$, we have $U^\prime\overline{\rho}_0=\int U^\prime U\rho_0U^\dagger dU=\int U\rho_0U^\dagger d(U^{\prime\dagger}U)U^\prime=\int U\rho_0U^\dagger dUU^\prime=\overline{\rho}_0U^\prime$, where we have used the translation-invariant property of the Haar measure, i.e., $d(U^{\prime\dagger}U)=dU$. It means that $\overline{\rho}_0$ commutes with all unitary operators in $\mathcal{A}_{\rho_0}^\prime$. Since every element of $\mathcal{A}_{\rho_0}^\prime$ can be written as a linear combination of unitary elements of $\mathcal{A}_{\rho_0}^\prime$, we have that $\overline{\rho}_0$ commutes with all elements of $\mathcal{A}_{\rho_0}^\prime$. Hence, $\overline{\rho}_0\in\mathcal{A}_{\rho_0}$. This completes the proof of Lemma \ref{lemma2}.

With the aid of Lemma \ref{lemma2}, we may now derive the structure of the SSM supported on $P_{\rho_0}\mathcal{H}$. According to the standard structure theorems for C*-algebras \cite{Davidson1996}, the algebra $\mathcal{A}_{\rho_0}$ and its commutant $\mathcal{A}_{\rho_0}^\prime$ have the matrix representations in a proper basis,
\begin{eqnarray}
\mathcal{A}_{\rho_0} \cong \bigoplus_\alpha \openone_{n_\alpha}\otimes \textrm{Mat}_{d_\alpha}(\CC), \label{sst1}
\end{eqnarray}
\begin{eqnarray}
\mathcal{A}_{\rho_0}^\prime \cong \bigoplus_\alpha\textrm{Mat}_{n_\alpha}(\CC)\otimes\openone_{d_\alpha},\label{sst2}
\end{eqnarray}
where $\alpha$ labels the $\alpha$-th irreducible representation of $\mathcal{A}_{\rho_0}$ with dimension $d_\alpha$ and multiplicity $n_\alpha$, $\openone_{n}$ ($n=n_\alpha,d_\alpha$) denotes the $n\times n$ identity matrix, and $\textrm{Mat}_{n}(\CC)$ ($n=n_\alpha,d_\alpha$) denotes the set of $n\times n$ matrices with complex entries.  Equations (\ref{sst1}) and (\ref{sst2}) indicate that $P_{\rho_0}\mathcal{H}$ has the decomposition \cite{Zanardi2001,Zanardi2004},
\begin{eqnarray}\label{sst3}
P_{\rho_0}\mathcal{H}=\bigoplus_\alpha\mathcal{H}_{\alpha,1}\otimes\mathcal{H}_{\alpha,2},
\end{eqnarray}
and correspondingly $\mathcal{A}_{\rho_0}$ and $\mathcal{A}_{\rho_0}^\prime$ have the structures,
\begin{eqnarray}
\mathcal{A}_{\rho_0}=\bigoplus_\alpha I_{\alpha,1}\otimes L(\mathcal{H}_{\alpha,2}),\label{sst4}
\end{eqnarray}
\begin{eqnarray}
\mathcal{A}_{\rho_0}^\prime&=&\bigoplus_\alpha L(\mathcal{H}_{\alpha,1})\otimes I_{\alpha,2},\label{sst5}
\end{eqnarray}
where $\mathcal{H}_{\alpha,1}$ and $\mathcal{H}_{\alpha,2}$ are the subspaces with dimensions $n_\alpha$ and $d_\alpha$, respectively,  and $I_{\alpha,1}$ and $I_{\alpha,2}$ are the identity operators on $\mathcal{H}_{\alpha,1}$ and $\mathcal{H}_{\alpha,2}$, respectively.

Note that the steady state that can induce the same SSM $\rho_0^{\frac{1}{2}}\mathcal{A}_{\rho_0}^\prime\rho_0^{\frac{1}{2}}$ is not unique. All the steady states that have the same support as $\rho_0$ lead  to the same SSM, i.e., $\rho_{10}^{\frac{1}{2}}\mathcal{A}_{\rho_{10}}^\prime\rho_{10}^{\frac{1}{2}}=\rho_{20}^{\frac{1}{2}}\mathcal{A}_{\rho_{20}}^\prime\rho_{20}^{\frac{1}{2}}$ as long as $\rho_{10}$ and $\rho_{20}$ have the same support. On the other hand, as we have proved in Lemma \ref{lemma2}, among the steady states with the same support, there is one steady state, $\overline{\rho}_0$, belonging to the modified noise algebra $\mathcal{A}_{\rho_0}$. We can determine the structure of the SSM supported on $P_{\rho_0}\mathcal{H}$ by the aid of $\overline{\rho}_0$, as $\rho_0$ and $\overline{\rho}_0$ correspond to the same SSM. Since $\overline{\rho}_0$ belongs to the modified noise algebra, it follows from Eq. (\ref{sst4}) that there exist density operators $\rho_{\alpha,2}$ on $\mathcal{H}_{\alpha,2}$ such that $\overline{\rho}_0=\sum_\alpha c_\alpha I_{\alpha,1}\otimes\rho_{\alpha,2}$ ($c_\alpha$ is the normalization constant). Using Eq. (\ref{sst5}) and Theorem \ref{th1}, we deduce that the SSM supported on $P_{\rho_0}\mathcal{H}$ is equal to $\bigoplus_\alpha L(\mathcal{H}_{\alpha,1})\otimes \rho_{\alpha,2}$. Hence, we arrive at the following structure theorem.

\begin{theorem}\label{th2}
The steady-state manifold supported on $P_{\rho_0}\mathcal{H}$ has the structure $\bigoplus_\alpha L(\mathcal{H}_{\alpha,1})\otimes \rho_{\alpha,2}$, where $L(\mathcal{H}_{\alpha,1})$ is the set of linear operators on $\mathcal{H}_{\alpha,1}$,  $\rho_{\alpha,2}$ is a fixed density operator on $\mathcal{H}_{\alpha,2}$, and $\mathcal{H}_{\alpha,1}$ and $\mathcal{H}_{\alpha,2}$ are defined in Eq. (\ref{sst3}).
\end{theorem}

Theorem \ref{th2}, describing the structure of SSMs, is applicable to both Markovian and non-Markovian systems. As a special case, when it is applied to Markovian systems, it gives the same result on the structure of steady states as presented in Ref. \cite{Baumartner}.

\section{Approach to find steady-state manifolds}\label{sec4}

Equation (\ref{sst4}) implies that any Hermitian operator in $\mathcal{A}_{\rho_0}$ can be generally expressed as $\sum_{\alpha j}\mu_{\alpha j}I_{\alpha,1}\otimes\ket{j}_{\alpha,2}\bra{j}$, where $\mu_{\alpha j}$  are real numbers and $\ket{j}_{\alpha,2}$ are pure states on $\mathcal{H}_{\alpha,2}$. We use $P$ to represent one of the Hermitian operators that are with nondegenerate spectra $\mu_{\alpha j}$, i.e., $P=\sum_{\alpha j}\mu_{\alpha j}I_{\alpha,1}\otimes\ket{j}_{\alpha,2}\bra{j}$ ~($\mu_{\alpha j}\neq\mu_{\alpha^\prime j^\prime}$ for $\alpha,j\neq\alpha^\prime,j^\prime$), of which the spectral projections are $I_{\alpha,1}\otimes\ket{j}_{\alpha,2}\bra{j}$.
Similarly, Eq. (\ref{sst5}) implies that any Hermitian operator in $\mathcal{A}_{\rho_0}^\prime$ can be generally expressed as $\sum_{\alpha i}\nu_{\alpha i}\ket{i}_{\alpha,1}\bra{i}\otimes I_{\alpha,2}$ with real numbers $\nu_{\alpha i}$ and pure states $\ket{i}_{\alpha,1}$ on $\mathcal{H}_{\alpha,1}$. We use $Q$ to represent one of the Hermitian operators that are with nondegenerate spectra $\nu_{\alpha j}$, i.e., $Q=\sum_{\alpha i}\nu_{\alpha i}\ket{i}_{\alpha,1}\bra{i}\otimes I_{\alpha,2}$ ~($\nu_{\alpha i}\neq\nu_{\alpha^\prime i^\prime}$ for $\alpha,i\neq\alpha^\prime,i^\prime$), of which the spectral projections are $\ket{i}_{\alpha,1}\bra{i}\otimes I_{\alpha,2}$. Certainly,  $P\in\mathcal{A}_{\rho_0}$ and  $Q\in\mathcal{A}_{\rho_0}^\prime$. Hereafter, we refer to such Hermitian operators that are with nondegenerate spectra as   spectrum-nondegenerate Hermitian operators for convenience.

To find the SSM on $P_{\rho_0}\mathcal{H}$, it is essential to identify the decomposition expressed by Eq. (\ref{sst3}), i.e., to determine the basis in which $P_{\rho_0}\mathcal{H}$ can be decomposed as $\bigoplus_\alpha\mathcal{H}_{\alpha,1}\otimes\mathcal{H}_{\alpha,2}$. We find that the basis can be obtained by choosing a pair of operators $P\in \mathcal{A}_{\rho_0}$ and $Q\in\mathcal{A}_{\rho_0}^\prime$, defined as above. By resorting to the spectral projections of $P$ and $Q$, i.e., $I_{\alpha,1}\otimes\ket{j}_{\alpha,2}\bra{j}$ and $\ket{i}_{\alpha,1}\bra{i}\otimes I_{\alpha,2}$, respectively, we can easily obtain the basis that corresponds to the decomposition (\ref{sst3}). In fact, the product of $I_{\alpha,1}\otimes\ket{j}_{\alpha,2}\bra{j}$ and $\ket{i}_{\alpha,1}\bra{i}\otimes I_{\alpha,2}$ gives $\ket{i}_{\alpha,1}\bra{i}\otimes\ket{j}_{\alpha,2}\bra{j}$, from which the states $\ket{i}_{\alpha,1}\otimes\ket{j}_{\alpha,2}$ can be derived. $\ket{i}_{\alpha,1}\otimes\ket{j}_{\alpha,2}$ can be taken as the basis of the decomposition in Eq. (\ref{sst3}). Therefore, to find a SSM based on a given steady state $\rho_0$, one may first construct the modified noise algebra $\mathcal{A}_{\rho_0}$ and its commutant $\mathcal{A}_{\rho_0}^\prime$, which can be realized by giving the generators of the algebras; then pick out two spectrum-nondegenerate Hermitian operators $P$ and $Q$ from $\mathcal{A}_{\rho_0}$ and $\mathcal{A}_{\rho_0}^\prime$, respectively; and finally identify the basis $\ket{i}_{\alpha,1}\otimes\ket{j}_{\alpha,2}$ by resorting to the spectral projections of $P$ and $Q$.

With the above analysis, we may now specify the approach to find SSM of an open system based on a given steady state $\rho_0$.

The first step is to construct the generating set of $\mathcal{A}_{\rho_0}$ and that of $\mathcal{A}_{\rho_0}^\prime$. Here, a generating set of an algebra means a subset of elements from which every element of the algebra can be expressed by a linear combination of products of the generators.  Obviously, the generating set of $\mathcal{A}_{\rho_0}$, denoted by $\{A_1,\dots,A_m\}$, can be simply taken as $\{\tilde{E}_k(t),~ \tilde{E}_k^\dagger(t)\}$ up to multiplicative scalars, which is just the definition of the modified noise algebra.  Then, the generating set of $\mathcal{A}_{\rho_0}^\prime$ can be obtained with the aid of the $\mathcal{A}_{\rho_0}$'s generating set. By definition, an operator $B$ belongs to $\mathcal{A}_{\rho_0}^\prime$ if and only if $[A_i,B]=0$ for all $i$, where $A_i\in\{\tilde{E}_k(t),~ \tilde{E}_k^\dagger(t)\}$. Note that any $m\times n$ matrix $B$ can be converted into a column vector $\textrm{vec}(B)=[B_{11},\dots,B_{m1},B_{12},\dots,B_{m2},\dots,B_{1n},\dots,B_{mn}]^T$ with the elements of the vector being a rearrangement of the elements of the matrix.  By vectorizing the equation $[A_i,B]=0$ and using the relations $\textrm{vec}(A_iBI)=(I\otimes A_i)\textrm{vec}(B)$ and $\textrm{vec}(IBA_i)=(A^T_i\otimes I)\textrm{vec}(B)$, we have $(I\otimes A_i-A_i^T\otimes I)\textrm{vec}(B)=0$. In this way, solving the equations $[A_i,B]=0$ is converted into solving the linear equations $(I\otimes A_i-A_i^T\otimes I)\textrm{vec}(B)=0$, for which there have been many standard methods. By solving these linear equations, we may obtain a complete set of linearly independent solutions of $\textrm{vec}(B)$, denoted as $\{\textrm{vec}(B_1),\dots,\textrm{vec}(B_n)\}$. Then, $\{B_1,\dots,B_n\}$, obtained by converting each vector $\textrm{vec}(B_i)$ back into matrix $B_i$, gives the generating set of $\mathcal{A}_{\rho_0}^\prime$.

The second step is to pick out two spectrum-nondegenerate Hermitian operators $P\in\mathcal{A}_{\rho_0}$ and $Q\in\mathcal{A}_{\rho_0}^\prime$. This can be realized by directly examining some pairs of Hermitian operators in $\mathcal{A}_{\rho_0}$ and $\mathcal{A}_{\rho_0}^\prime$. Let $P\in\mathcal{A}_{\rho_0}$ and $Q\in\mathcal{A}_{\rho_0}^\prime$ be Hermitian operators, and $P_j$ and $Q_i$ be their spectral projections, i.e., $P=\sum_j\mu_jP_j$ and $Q=\sum_i\nu_iQ_i$, where $\mu_j$ and $\nu_i$ are the spectra corresponding to $P_j$ and $Q_i$, respectively.  Whether $P$ and $Q$ are with nondegenerate spectra can be examined by resorting to the generating sets of $\mathcal{A}_{\rho_0}$ and $\mathcal{A}_{\rho_0}^\prime$. In fact, Hermitian operators $P$ and $Q$ are with nondegenerate spectra if and only if all their spectral projections $P_j$ and $Q_i$ satisfy respectively
\begin{eqnarray}
P_jA_1P_j\propto P_j,~~~ P_jA_2P_j\propto P_j,~\dots,~~P_jA_mP_j\propto P_j,\label{proj1}
\end{eqnarray}
and
\begin{eqnarray}
Q_iB_1Q_i\propto Q_i,~~~Q_iB_2Q_i\propto Q_i,~\dots,~~Q_iB_nQ_i\propto Q_i, \label{proj2}
\end{eqnarray}
where $X\propto Y$ means that $X=\lambda Y$ with $\lambda$ being a complex number \cite{Holbrook2003}. Therefore, to find a pair of spectrum-nondegenerate Hermitian operators, one can first choose two Hermitian operators $P\in\mathcal{A}_{\rho_0}$ and $Q\in\mathcal{A}_{\rho_0}^\prime$, work out their spectral projections, and then check whether they match the criterion expressed by Eqs. (\ref{proj1}) and (\ref{proj2}). If not, one can repeat this procedure until the criterion is matched. Since the family of Hermitian operators with degenerate spectra is only a subset of measure zero and hence most of Hermitian operators in $\mathcal{A}_{\rho_0}$ and $\mathcal{A}_{\rho_0}^\prime$ are with nondegenerate spectra, it is easy to pick out two desired operators $P$ and $Q$.

The third step is to identify the basis of the decomposition (\ref{sst3}) by using the spectral projections of $P$ and $Q$, and to further obtain the SSM, i.e., the density operators with the structure $\bigoplus_\alpha L(\mathcal{H}_{\alpha,1})\otimes\rho_{\alpha,2}$.  To this end, we subdivide the set of spectral projections $\{P_j,Q_i,~j=1,2,\cdots,\sum_\alpha d_\alpha,~i=1,2,\cdots,\sum_\alpha n_\alpha\}$ into subsets. $P_j$ and $Q_i$ belong to the same subset if and only if $Q_iP_j\neq 0$. It is easy to check whether some projections belong to a same set by straightforwardly multiplying each other. We renumber the spectral projections belonging to the $\alpha$-th subset and denote them by $P_{j}^{(\alpha)}$ and $Q_{i}^{(\alpha)}$. It is interesting to note that each subset, $\{P_{j}^{(\alpha)},Q_{i}^{(\alpha)},~j=1,\cdots,d_\alpha,~i=1,\cdots,n_\alpha\}$, just corresponds to an irreducible representation subspace in Eq. (\ref{sst3}), where $d_\alpha$ and $n_\alpha$ are respectively the total numbers of $P_j$ and $Q_i$ in the $\alpha$-th subset. To obtain the basis of the $\alpha$-th subspace, we only need to calculate the eigenvectors of the products $Q_{i}^{(\alpha)}P_{j}^{(\alpha)}$ with eigenvalue $1$, denoted as $\ket{i,j}_\alpha$.
Then, $\{\ket{i,j}_\alpha, ~ i=i(\alpha)=1,\cdots,n_\alpha,~j=j(\alpha)=1,\cdots,d_\alpha\}$ forms the basis of the $\alpha$-th subspace, and all the eigenvectors $\{\ket{i,j}_\alpha\}$ (for all $\alpha$) form the basis of the decomposition expressed by Eq. (\ref{sst3}). With this basis, we can easily find the SSM. Indeed, by expressing $\rho_0$ in the basis $\ket{i,j}_\alpha$, we can work out $\rho_{\alpha,2}$ appearing in Theorem \ref{th2}, with which we can immediately write out all the steady states supported on $P_{\rho_0}\mathcal{H}$, i.e., the density operators belonging to $\bigoplus_\alpha L(\mathcal{H}_{\alpha,1})\otimes\rho_{\alpha,2}$.

\section{Examples}\label{sec5}

So far, we have put forward an approach to find the SSM based on a given steady state $\rho_0$. When $\rho_0$ is the steady state with the maximum support, the SSM obtained by our approach will be the whole SSM of the system. When $\rho_0$ is a steady state but not the one with the maximum support, the SSM obtained by our approach is a subset of the whole SSM, i.e., the SSM supported on $P_{\rho_0}\mathcal{H}$.  It is interesting to note that $\rho_0=cI$ is a steady state with the maximum support for the systems under unital CPTP maps, and the SSM obtained by our approach contains all the steady states for these systems. Our approach is applicable to both Markovian and non-Markovian systems.  We now give three simple examples to illustrate the usefulness of our approach.

{\textit Example 1.}~~   We first apply our approach to a Markovian system.  Consider the well-known model, the open system of multiple qubits experiencing collective decoherence~\cite{Zanardi1997}. For ease of notation, we take the 3-qubit case as an example. The dynamics of the 3-qubit system experiencing collective decoherence, in the Markovian approximation, is governed by the Lindblad equation with the following Liouvillian \cite{Zanardi2014},
\begin{eqnarray}
\mathcal{L}(\rho)=\sum_{k=x,y,z}\gamma_k\left(S_k\rho S_k^\dagger-\frac{1}{2}\{S_k^\dagger S_k,\rho\}\right),
\end{eqnarray}
where $S_k=\sum_{i=1}^3 \sigma_k^i$ are collective spin operators with $\sigma_k^i$ being the Pauli operator for the $i$-th qubit.

For this model, it is easy to check that the identity operator is a steady state, and therefore we can take $\rho_0=cI$, which is with the maximum support. By directly following our approach, we can obtain the SSM of this system.

First, we construct the generating sets of  $\mathcal{A}_{\rho_0}$ and $\mathcal{A}_{\rho_0}^\prime$. The modified noise algebra $\mathcal{A}_{\rho_0}$ and its commutant $\mathcal{A}_{\rho_0}^\prime$ are just equal to the noise algebra $\mathcal{A}$ and the commutant $\mathcal{A}^\prime$, respectively, due to $\rho_0=cI$.  The generating set of $\mathcal{A}$ can be taken as $\{S_1,  S_2, S_3\}$. By resolving linear equations $(I\otimes S_k-S_k^T\otimes I)\textrm{vec}(B)=0$, $k=x,y,z$, we have the generating set of $\mathcal{A}^\prime$, $\{B_1, B_2, B_3\}=\{\boldsymbol{\sigma}_1\cdot\boldsymbol{\sigma}_2,~\boldsymbol{\sigma}_2\cdot\boldsymbol{\sigma}_3,
~\boldsymbol{\sigma}_1\cdot\boldsymbol{\sigma}_3\}$, where $\boldsymbol{\sigma}_i=(\sigma_x^i,\sigma_y^i,\sigma_z^i)$.

Second, we pick out two spectrum-nondegenerate Hermitian operators $P\in\mathcal{A}_{\rho_0}$ and $Q\in\mathcal{A}_{\rho_0}^\prime$.  We choose $P=S_x^2+S_y^2+S_z^2+S_z$ and $Q=\boldsymbol{\sigma}_1\cdot\boldsymbol{\sigma}_3+\boldsymbol{\sigma}_2\cdot\boldsymbol{\sigma}_3$. The spectral projections of $P$ read
$P_1^{(1)}=\ket{\psi_0}\bra{\psi_0}+\ket{\psi_2}\bra{\psi_2}$,~ $P_2^{(1)}=\ket{\psi_1}\bra{\psi_1}+\ket{\psi_3}\bra{\psi_3}$, ~$P_1^{(2)}=\ket{\psi_4}\bra{\psi_4}$, ~$P_2^{(2)}=\ket{\psi_5}\bra{\psi_5}$, ~$P_3^{(2)}=\ket{\psi_6}\bra{\psi_6}$,
~$P_4^{(2)}=\ket{\psi_7}\bra{\psi_7}$,
and the spectral projections of $Q$ read
$Q_1^{(1)}=\ket{\psi_0}\bra{\psi_0}+\ket{\psi_1}\bra{\psi_1}$, ~$Q_2^{(1)}=\ket{\psi_2}\bra{\psi_2}+\ket{\psi_3}\bra{\psi_3}$,  ~$Q_1^{(2)}=\ket{\psi_4}\bra{\psi_4}+\ket{\psi_5}\bra{\psi_5}+\ket{\psi_6}\bra{\psi_6}+\ket{\psi_7}\bra{\psi_7}$,
where $\ket{\psi_0}=(\ket{010}-\ket{100})/\sqrt{2}$,
~$\ket{\psi_1}=(\ket{011}-\ket{101})/\sqrt{2}$,
~$\ket{\psi_2}=(2\ket{001}-\ket{010}-\ket{100})/\sqrt{6}$,
~$\ket{\psi_3}=(-2\ket{110}+\ket{011}+\ket{101})/\sqrt{6}$,
~$\ket{\psi_4}=\ket{000}$,
~$\ket{\psi_5}=(\ket{001}+\ket{010}+\ket{100})/\sqrt{3}$,
~$\ket{\psi_6}=(\ket{110}+\ket{011}+\ket{101})/\sqrt{3}$, and
$\ket{\psi_7}=\ket{111}$.
By substituting these spectral projections into Eqs. (\ref{proj1}) and (\ref{proj2}), it is easy to verify that they satisfy the criterion in Eqs. (\ref{proj1}) and (\ref{proj2}), and therefore $P$ and $Q$ are two  spectrum-nondegenerate Hermitian operators.

Third, we identify the basis of the decomposition expressed as Eq. (\ref{sst3}) by using the spectral projections of $P$ and $Q$, and further obtain the SSM.
By multiplying the spectral projections by each other and checking whether their products are nonzero, the set of spectral projections are subdivided  into two subsets, $\{P_1^{(1)},P_2^{(1)},Q_1^{(1)},Q_2^{(1)}\}$ and $\{P_1^{(2)},P_2^{(2)},P_3^{(2)},P_4^{(2)},Q_1^{(2)}\}$. Note that the $\alpha$-th subset corresponds to the $\alpha$-th irreducible representation, the number of $P_j^{(\alpha)}$ is equal to the dimension $d_\alpha$, and the number of $Q_i^{(\alpha)}$ is equal to the multiplicity $n_\alpha$. It follows that there are two irreducible representations in Eq. (\ref{sst3}), i.e., $\alpha=1,2$, the first irreducible representation is with dimension $2$ and multiplicity 2, and the second irreducible representation is with dimension 4 and multiplicity 1. Then, by computing the eigenvectors corresponding to the eigenvalue $1$, we obtain that the eigenvectors of $Q_1^{(1)}P_1^{(1)}$, $Q_1^{(1)}P_2^{(1)}$, $Q_2^{(1)}P_1^{(1)}$, $Q_2^{(1)}P_2^{(1)}$, $Q_1^{(2)}P_1^{(2)}$, $Q_1^{(2)}P_2^{(2)}$, $Q_1^{(2)}P_3^{(2)}$, and $Q_1^{(2)}P_4^{(2)}$ are $\ket{\psi_0}$, $\ket{\psi_1},\ket{\psi_2}$,
$\ket{\psi_3}$, $\ket{\psi_4}$, $\ket{\psi_5}$, $\ket{\psi_6}$, and $\ket{\psi_7}$, respectively. These eigenvectors form a basis for the decomposition in Eq. (\ref{sst3}). With this basis, we can deduce from Theorem \ref{th2} that the SSM is with the structure $\left(\textrm{Mat}_2(\CC)\otimes\openone_2\right)\oplus\openone_4$, implying that the steady states are the density operators with the matrix representatives belonging to $\left(\textrm{Mat}_2(\CC)\otimes\openone_2\right)\oplus\openone_4$ in this basis.

{\textit Example 2.}~~   We now apply our approach to a non-Markovian system. Consider a simple error model, a 2-qubit system under the dynamical map,
\begin{eqnarray}\label{ex1}
\mathcal{E}_t(\rho)=\left(1-\int_0^tf(\mu)d\mu\right)\rho+\int_0^tf(\mu)d\mu~\mathcal{P}(\rho),
\end{eqnarray}
where $\mathcal{P}(\rho):=\sum_{k=0}^3E_k\rho E_k^\dagger$ is a CPTP map with Kraus operators $E_0=\frac{1}{2}I\otimes I$, $E_1=\frac{1}{2}\sigma_x\otimes I$, $E_2=\frac{1}{2}\sigma_y\otimes\sigma_z$, and $E_3=\frac{1}{2}\sigma_z\otimes\sigma_z$, and $f(t)$ is a real function satisfying $0\leq\int_0^tf(\mu)d\mu\leq 1$ for any $t>0$ and $f(t)<0$ for some time interval \cite{Chruscinski2010}. Equation (\ref{ex1}) represents non-Markovian dynamics. This can be seen by converting Eq. (\ref{ex1}) into a non-Markovian master equation, $\partial_t\rho=\alpha(t)\mathcal{L}(\rho)$, with $\alpha(t)=f(t)/(1-\int_0^tf(\mu)d\mu)$ and $\mathcal{L}=\mathcal{P}-\mathcal{I}$, where $\mathcal{I}$ is the identity map \cite{Chruscinski2010}.

By taking $\rho_0=cI$ and following our approach, the SSM of the non-Markovian system can be obtained.

First, we construct the generating sets of $\mathcal{A}_{\rho_0}$ and $\mathcal{A}_{\rho_0}^\prime$. For $\rho_0=cI$, there is $\mathcal{A}_{\rho_0}=\mathcal{A}$ and $\mathcal{A}_{\rho_0}^\prime=\mathcal{A}^\prime$.
The generating set of $\mathcal{A}$ can be taken as $\{E_i,i=1,2,3\}$, and by resolving linear equations $(I\otimes E_i-E_i^T\otimes I)\textrm{vec}(B)=0$, $i=1,2,3$, we have the generating set of $\mathcal{A}^\prime$,  $\{B_1, B_2, B_3\}=\{I\otimes\sigma_z,\sigma_x\otimes\sigma_y,\sigma_x\otimes\sigma_x\}$.

Second, we pick out two spectrum-nondegenerate Hermitian operators $P\in\mathcal{A}_{\rho_0}$ and $Q\in\mathcal{A}_{\rho_0}^\prime$.  We choose $P=E_3$ and $Q=\sigma_x\otimes\sigma_x$.  The spectral projections of $P$ read
$P_1=\ket{00}\bra{00}+\ket{11}\bra{11}$, $P_2=\ket{01}\bra{01}+\ket{10}\bra{10}$,
and the spectral projections of $Q$ read
$Q_1=\ket{++}\bra{++}+\ket{--}\bra{--}$, and $Q_2=\ket{+-}\bra{+-}+\ket{-+}\bra{-+}$, where $\ket{\pm}:=(\ket{0}\pm\ket{1})/\sqrt{2}$.
By substituting these spectral projections into Eqs. (\ref{proj1}) and (\ref{proj2}), it is easy to verify that they satisfy the criterion expressed by these equations, and therefore $P$ and $Q$ are two spectrum-nondegenerate Hermitian operators.

Third, we identify the basis of the decomposition expressed by Eq. (\ref{sst3}) by using the spectral projections of $P$ and $Q$, and obtain the SSM. By multiplying the spectral projections by each other and checking whether their products are nonzero, we find that $Q_iP_j\neq0$ for all $i,j=1,2$. It means that all the spectral projections belong to one set, $\{P_1,P_2,Q_1,Q_2\}$.
Note that the $\alpha$-th subset corresponds to the $\alpha$-th irreducible representation, the number of $P_j^{(\alpha)}$ is equal to the dimension $d_\alpha$, and the number of $Q_i^{(\alpha)}$ is equal to the multiplicity $n_\alpha$. It follows that there is only one irreducible representation in Eq. (\ref{sst3}) with dimension 2 and multiplicity 2. Then, by computing the eigenvectors corresponding to the eigenvalue 1, we obtain that the eigenvectors of $Q_1P_1$, $Q_1P_2$, $Q_2P_1$, and $Q_2P_2$ are $(\ket{00}+\ket{11})/\sqrt{2}$, $(\ket{10}+\ket{10})/\sqrt{2}$, $(\ket{00}-\ket{11})/\sqrt{2}$, and $(\ket{10}-\ket{01})/\sqrt{2}$, respectively. These eigenvectors form a basis for the induced decomposition in Eq. (\ref{sst3}). With this basis, we can deduce from Theorem \ref{th2} that the SSM  is with the structure $\textrm{Mat}_2(\CC)\otimes\openone_2$, implying that the steady states are the density operators with the matrix representatives belonging to $\textrm{Mat}_2(\CC)\otimes\openone_2$ in this basis.

\textit{Example 3.} We then apply our approach to an open system under a non-unital CPTP map. Consider the error model, a 3-qubit system under the CPTP map,
\begin{eqnarray}
\mathcal{E}_t(\rho)=\sum_{k=0}^{3}E_k\rho E_k^\dagger,
\end{eqnarray}
where $E_0=\sqrt{1-2p}I\otimes I\otimes I$, $E_1=\sqrt{p}\ket{0}\bra{0}\otimes\sigma_x\otimes\sigma_x$, $E_2=\sqrt{p}\ket{0}\bra{0}\otimes\sigma_z\otimes I$, and $E_3=\sqrt{2p}\ket{0}\bra{1}\otimes I\otimes I$, with $0\leq p\leq 1/2$ being a parameter dependent on time $t$.

For this model, it is easy to check that $\rho_0=\frac{1}{4}\ket{0}\bra{0}\otimes I\otimes I$ is a steady state. By directly following our approach, we can obtain the SSM supported on $P_{\rho_0}\mathcal{H}$, where $P_{\rho_0}=\ket{0}\bra{0}\otimes I\otimes I$.

First, we construct the generating sets of $\mathcal{A}_{\rho_0}$ and $\mathcal{A}_{\rho_0}^\prime$. Direct calculations show that there are three modified Kraus operators, namely, $\tilde{E}_0=\sqrt{1-2p}\ket{0}\bra{0}\otimes I\otimes I$, $\tilde{E}_1=\sqrt{p}\ket{0}\bra{0}\otimes\sigma_x\otimes\sigma_x$, and $\tilde{E}_2=\sqrt{p}\ket{0}\bra{0}\otimes\sigma_z\otimes I$. Hence, the generating set of $\mathcal{A}_{\rho_0}$ can be taken as $\{\tilde{E}_i,i=0,1,2\}$. By resolving linear equations $(I\otimes \tilde{E}_i-\tilde{E}_i^T\otimes I)\textrm{vec}(B)=0$, $i=0,1,2$, we have the generating set of $\mathcal{A}_{\rho_0}^\prime$,  $\{B_1, B_2, B_3\}=\{\ket{0}\bra{0}\otimes I\otimes\sigma_x, \ket{0}\bra{0}\otimes \sigma_z\otimes\sigma_y, \ket{0}\bra{0}\otimes \sigma_z\otimes\sigma_z\}$.

Second, we pick out two spectrum-nondegenerate Hermitian operators $P\in\mathcal{A}_{\rho_0}$ and $Q\in\mathcal{A}_{\rho_0}^\prime$.  We choose $P=\tilde{E}_2$ and $Q=\ket{0}\bra{0}\otimes I\otimes\sigma_x$.  The spectral projections of $P$ read
$P_1=\ket{0}\bra{0}\otimes\ket{0}\bra{0}\otimes I$, $P_2=\ket{0}\bra{0}\otimes\ket{1}\bra{1}\otimes I$,
and the spectral projections of $Q$ read
$Q_1=\ket{0}\bra{0}\otimes I\otimes\ket{+}\bra{+}$, and $Q_2=\ket{0}\bra{0}\otimes I\otimes\ket{-}\bra{-}$.
By substituting these spectral projections into Eqs. (\ref{proj1}) and (\ref{proj2}), it is easy to verify that they satisfy the criterion in Eqs. (\ref{proj1}) and (\ref{proj2}), and therefore $P$ and $Q$ are two spectrum-nondegenerate Hermitian operators.

Third, we identify the basis of the decomposition expressed by Eq. (\ref{sst3}) by using the spectral projections of $P$ and $Q$, and obtain the SSM supported on $P_{\rho_0}\mathcal{H}$. By multiplying the spectral projections by each other and checking whether their products are nonzero, we find that $Q_iP_j\neq0$ for all $i,j=1,2$. It means that all the spectral projections belong to one set, $\{P_1,P_2,Q_1,Q_2\}$. Hence, there is only one irreducible representation in Eq. (\ref{sst3}) with dimension 2 and multiplicity 2. Then, by computing the eigenvectors corresponding to the eigenvalue 1, we obtain that the eigenvectors of $Q_1P_1$, $Q_1P_2$, $Q_2P_1$, and $Q_2P_2$ are $\ket{00+}$, $\ket{01+}$, $\ket{00-}$, and $\ket{01-}$, respectively. These eigenvectors form a basis for the decomposition in Eq. (\ref{sst3}). With this basis, we can deduce from Theorem \ref{th2} that the SSM supported on $P_{\rho_0}\mathcal{H}$ is with the structure $\textrm{Mat}_2(\CC)\otimes\openone_2$, implying that the steady states are the density operators with the matrix representatives belonging to $\textrm{Mat}_2(\CC)\otimes\openone_2$ in this basis.

\section{Summary and remarks}\label{sec6}

In summary, we have proposed an approach for finding steady-state manifolds of open quantum systems. The proposed approach can be briefly summarized as the three steps: i) Construct the generating sets of the modified noise algebra and its commutant. ii) Pick out a pair of spectrum-nondegenerate Hermitian operators from the modified noise algebra and its commutant, respectively, by resorting to the generating sets. iii) Identify the basis of the decomposition in Eq. (\ref{sst3}) to obtain the steady-state manifold, by using the spectral projections of the pair of operators. Compared with the previous works \cite{Diehl2010,Lee2013,Jin2013,Weimer2015,Cui2015}, which are for Markovian systems, our approach is applicable to general open systems, both Markovian and non-Markovian systems, and can help to determine the structure of steady-state manifolds. Three examples are presented to illustrate the applications of our approach.

It is worth noting that our approach is based on an arbitrarily given steady state $\rho_0$. With the aid of one given state $\rho_0$, the steady-state manifold on the support subspace of the given state can be obtained by simply following the standard steps of the approach. In particular, when $\rho_0$ is a steady state with the maximum support, the steady-state manifold obtained by our approach will contain all the steady states of the system. Our work reduces the problem of finding a manifold of steady states to that of finding only one steady state, which is indeed an interesting progress towards completely solving the difficult problem. For finding a steady state $\rho_0$, there have been some known results. For instance, $\rho_0=cI$ is always a steady state for open systems under unital dynamical maps, and for open systems in which the external time-dependent fields are absent, $\rho_0$ may be taken as the Gibbs state. In general, one may find a steady state $\rho_0$ by using various methods, such as those in the previous papers \cite{Diehl2010,Lee2013,Jin2013,Weimer2015,Cui2015}.

Besides, we have introduced the notions of the modified noise algebra and its commutant, and proved two theorems on the structure of steady-state manifolds of general open systems, which themselves are interesting findings too.

\section*{Acknowledgments}

This work was supported by the China Postdoctoral Science Foundation under Grant No. 2016M592173. X.D.Y. acknowledges support from the National Natural Science Foundation
of China through Grant No. 11575101. H.L.H. acknowledges support from the National Natural Science Foundation
of China through Grant No. 11571199. D.M.T. acknowledges support from the National Basic Research Program of China through Grant No. 2015CB921004.

\appendix
\setcounter{equation}{0}

\section{}\label{appA}
Here, we prove that for all $\rho\in L(P_{\rho_0}\mathcal{H})$ and $U\in\mathcal{A}_{\rho_0}^\prime$,
\begin{eqnarray}\label{A1}
\mathcal{E}_t(U\rho U^\dagger)=U\mathcal{E}_t(\rho)U^\dagger,
\end{eqnarray}
which has been used in the proof of Theorem \ref{th2}.

We denote by $\textrm{M}$ the SSM supported on the subspace $P_{\rho_0}\mathcal{H}$, and by
$F(\mathcal{E}_t)$ and $F(\mathcal{E}_t^*)$ the set of fixed points of $\mathcal{E}_t$ and that of $\mathcal{E}_t^*$, respectively, $\textrm{M}:=\{X\in L(P_{\rho_0}\mathcal{H})|\mathcal{E}_t(X)=X,\forall t\}$, $F(\mathcal{E}_t):=\{X\in L(P_{\rho_0}\mathcal{H})|\mathcal{E}_t(X)=X\}$, and $F(\mathcal{E}_t^*):=\{X\in L(P_{\rho_0}\mathcal{H})|\mathcal{E}_t^*(X)=X\}$. We need to prove that $\mathcal{A}_{\rho_0}=\textrm{alg}\{E_k(t)|_{P_{\rho_0}\mathcal{H}},E_k(t)|_{P_{\rho_0}\mathcal{H}}^\dagger,\forall k,t\}$, where $E_k(t)|_{P_{\rho_0}\mathcal{H}}=P_{\rho_0}E_k(t)P_{\rho_0}$. This is done as follows.

First, we introduce an auxiliary map defined as
$\mathcal{P}_t:=\lim_{N\rightarrow\infty}\frac{1}{N}\sum_{n=1}^N\mathcal{E}_t^n$.
In Ref. \cite{Zhang2016}, it has been shown that $\mathcal{P}_t$ is a projection onto $F(\mathcal{E}_t)$, i.e., $\mathcal{P}_t[L(P_{\rho_0}\mathcal{H})]=F(\mathcal{E}_t)$. Following the same arguments as in Ref. \cite{Zhang2016}, we can further show that the dual map $\mathcal{P}_t^*$ is a projection onto $F(\mathcal{E}_t^*)$, i.e., $\mathcal{P}_t^*[L(P_{\rho_0}\mathcal{H})]=F(\mathcal{E}_t^*)$.

Second, with the aid of the auxiliary map, we establish the relation between $F(\mathcal{E}_t)$ and $F(\mathcal{E}_t^*)$.
In Ref. \cite{Zhang2016}, it has been shown that an explicit expression of the auxiliary map reads
$\mathcal{P}_t(X)=\sum_\alpha\Tr_{\alpha,1}(P_\alpha XP_{\alpha})\otimes\rho_{\alpha,2}$. Here, this expression corresponds to a decomposition of $P_{\rho_0}\mathcal{H}$, $P_{\rho_0}\mathcal{H}=\bigoplus_\alpha\mathcal{H}_{\alpha,1}\otimes\mathcal{H}_{\alpha,2}$, in which $P_\alpha$ denotes the orthogonal projector onto the subspace $\mathcal{H}_{\alpha,1}\otimes\mathcal{H}_{\alpha,2}$, and $\rho_{\alpha,2}$ is a fixed density operator on $\mathcal{H}_{\alpha,2}$.
From the expression of $\mathcal{P}_t$, we have that the explicit expression of $\mathcal{P}_t^*$ reads
$\mathcal{P}_t^*(X)=\sum_\alpha\Tr_{\alpha,1}(\rho_{\alpha,2} X)\otimes I_{\alpha,2}$.
It follows that $F(\mathcal{E}_t)=\bigoplus_\alpha L(\mathcal{H}_{\alpha,1})\otimes\rho_{\alpha,2}$ and $F(\mathcal{E}_t^*)=\bigoplus_\alpha L(\mathcal{H}_{\alpha,1})\otimes I_{\alpha,2}$. Besides, $\rho_0\in F(\mathcal{E}_t)$ and hence $\rho_0=\sum_\alpha X_{\alpha,1}\otimes\rho_{\alpha,2}$ for some positive operators $X_{\alpha,1}$. Therefore, we have
\begin{eqnarray}\label{relation}
F(\mathcal{E}_t)=\rho_0^{\frac{1}{2}}F(\mathcal{E}_t^*)\rho_0^{\frac{1}{2}}.
\end{eqnarray}

Third, we establish the relation between $F(\mathcal{E}_t^*)$ and the Kraus operators and further show that $\mathcal{A}_{\rho_0}=\textrm{alg}\{E_k(t)|_{P_{\rho_0}\mathcal{H}},E_k(t)|_{P_{\rho_0}\mathcal{H}}^\dagger,\forall k,t\}$.
Note that $\mathcal{E}_t^*$ is a unital CP map and its dual map $\mathcal{E}_t$ has a full-rank fixed point $\rho_0$. Following the same arguments as those in the proof of Lemma \ref{lemma1}, we have that
\begin{eqnarray}\label{A-F}
F(\mathcal{E}_t^*)=\textrm{alg}\{E_k(t)|_{P_{\rho_0}\mathcal{H}},E_k(t)|_{P_{\rho_0}\mathcal{H}}^\dagger,\forall k\}^\prime.
\end{eqnarray}
With the aid of Eqs. (\ref{relation}) and (\ref{A-F}) and noting that $\textrm{M}=\bigcap_tF(\mathcal{E}_t)$,  we have
\begin{eqnarray}\label{A-SF}
\textrm{M}&=&\bigcap_t F(\mathcal{E}_t)=\bigcap_t\rho_0^{\frac{1}{2}}F(\mathcal{E}_t^*)\rho_0^{\frac{1}{2}}\nonumber\\
&=&\bigcap_t\rho_0^{\frac{1}{2}}\textrm{alg}\{E_k(t)|_{P_{\rho_0}\mathcal{H}},E_k(t)|_{P_{\rho_0}\mathcal{H}}^\dagger,\forall k\}^\prime\rho_0^{\frac{1}{2}}\nonumber\\
&=&\rho_0^{\frac{1}{2}}\left(\bigcap_t\textrm{alg}\{E_k(t)|_{P_{\rho_0}\mathcal{H}},E_k(t)|_{P_{\rho_0}\mathcal{H}}^\dagger,\forall k\}^\prime\right)\rho_0^{\frac{1}{2}}\nonumber\\
&=&\rho_0^{\frac{1}{2}}\textrm{alg}\{E_k(t)|_{P_{\rho_0}\mathcal{H}},E_k(t)|_{P_{\rho_0}\mathcal{H}}^\dagger,\forall k,t\}^\prime\rho_0^{\frac{1}{2}}.
\end{eqnarray}
Comparing Eq. (\ref{A-SF}) with the expression of the SSM in Theorem \ref{th1}, we have
\begin{eqnarray}\label{S-alg}
\mathcal{A}_{\rho_0}=\textrm{alg}\{E_k(t)|_{P_{\rho_0}\mathcal{H}},E_k(t)|_{P_{\rho_0}\mathcal{H}}^\dagger,\forall k,t\}.
\end{eqnarray}

Equation (\ref{S-alg}) indicates that $[U,E_k(t)|_{P_{\rho_0}\mathcal{H}}]=0$ for $U\in\mathcal{A}_{\rho_0}^\prime$. We then obtain \begin{eqnarray}\label{end}
\mathcal{E}_t(U\rho U^\dagger)&=&\sum_k E_k(t)|_{P_{\rho_0}\mathcal{H}}U\rho U^\dagger E_k(t)|_{P_{\rho_0}\mathcal{H}}^\dagger\nonumber\\
&=&\sum_k UE_k(t)|_{P_{\rho_0}\mathcal{H}}\rho E_k(t)|_{P_{\rho_0}\mathcal{H}}^\dagger U^\dagger\nonumber\\
&=&U\mathcal{E}_t(\rho)U^\dagger.
\end{eqnarray}
This completes the proof of Eq. (\ref{A1}).


\begin{thebibliography}{99}
\bibitem{Zanardi1997} P. Zanardi and M. Rasetti, Phys. Rev. Lett. \textbf{79}, 3306 (1997).
\bibitem{Lidar1998} D. A. Lidar, I. L. Chuang, and K. B. Whaley, Phys. Rev. Lett.
   \textbf{81}, 2594 (1998).
\bibitem{Ollerenshaw2003} J. E. Ollerenshaw, D. A. Lidar, and L. E. Kay, Phys. Rev. Lett. \textbf{91}, 217904 (2003).
\bibitem{Mohseni2003} M. Mohseni, J. S. Lundeen, K. J. Resch, and A. M. Steinberg, Phys. Rev. Lett. \textbf{91}, 187903 (2003).
\bibitem{Viola2000} E. Knill, R. Laflamme, and L. Viola, Phys. Rev. Lett. \textbf{84}, 2525 (2000).
\bibitem{Zanardi2000} P. Zanardi, Phys. Rev. A \textbf{63}, 012301 (2000).
\bibitem{Viola2001} L. Viola, E. M. Fortunato, M. A. Pravia, E. Knill1, R. Laflamme, and D. G. Cory, Science \textbf{293}, 2059 (2001).
\bibitem{Diehl2008} S. Diehl, A. Micheli, A. Kantian, B. Kraus, H. P. B\"{u}chler, and P. Zoller, Nat. Phys. \textbf{4}, 878 (2008).
\bibitem{Kastoryano2011} M. J. Kastoryano, F. Reiter, and A. S. S{\o}rensen, Phys. Rev. Lett. \textbf{106}, 090502 (2011).
\bibitem{Krauter2011} H. Krauter, C. A. Muschik, K. Jensen, W. Wasilewski,
J. M. Petersen, J. I. Cirac, and E. S. Polzik, Phys. Rev. Lett. \textbf{107}, 080503 (2011).
\bibitem{Torre2013} E. G. Dalla Torre, J. Otterbach, E. Demler, V. Vuletic, and M. D. Lukin, Phys. Rev. Lett. \textbf{110}, 120402 (2013).
\bibitem{Carr2013} A. W. Carr and M. Saffman, Phys. Rev. Lett. \textbf{111}, 033607 (2013).
\bibitem{Znidari2016} M. \v{Z}nidari\v{c}, Phys. Rev. Lett. \textbf{116}, 030403 (2016).
\bibitem{Wu2005} L.-A. Wu, P. Zanardi, and D. A. Lidar, Phys. Rev. Lett. \textbf{95}, 130501 (2005).
\bibitem{Oreshkov2009} O. Oreshkov, Phys. Rev. Lett. \textbf{103}, 090502 (2009).
\bibitem{Xu2012} G. F. Xu, J. Zhang, D. M. Tong, E. Sj\"{o}qvist, and L. C. Kwek, Phys. Rev. Lett. \textbf{109}, 170501 (2012).
\bibitem{Feng2013} G. Feng, G. F. Xu, and G. Long, Phys. Rev. Lett. \textbf{110}, 190501 (2013).
\bibitem{Xu2014} G. F. Xu and G. Long, Sci. Rep. \textbf{4}, 6814 (2014).
\bibitem{ZhangJ2014} J. Zhang, L. C. Kwek, E. Sj\"{o}qvist, D. M. Tong, and P. Zanardi, Phys. Rev. A \textbf{89}, 042302 (2014).
\bibitem{Verstraete2009} F. Verstraete, M. M. Wolf, and J. I. Cirac, Nat. Phys. \textbf{5}, 633 (2009).
\bibitem{Zanardi2014} P. Zanardi and L. Campos Venuti, Phys. Rev. Lett. \textbf{113}, 240406 (2014).
\bibitem{Barreiro2011} J. T. Barreiro, M. M\"{u}ller, P. Schindler, D. Nigg, T. Monz, M. Chwalla, M, Hennrich, C. F. Roos, P. Zoller, and R. Blatt, Nature (London) \textbf{470}, 486 (2011).
\bibitem{Stannigel2014} K. Stannigel, P. Hauke, D. Marcos, M. Hafezi, S. Diehl, M. Dalmonte, and P. Zoller, Phys. Rev. Lett. \textbf{112}, 120406 (2014).
\bibitem{Vollbrecht2011} K. G. H. Vollbrecht, C. A. Muschik, and J. I. Cirac, Phys. Rev. Lett. \textbf{107}, 120502 (2011).
\bibitem{Mari2012} A. Mari and J. Eisert, Phys. Rev. Lett. \textbf{108}, 120602 (2012).
\bibitem{Diehl2010} S. Diehl, A. Tomadin, A. Micheli, R. Fazio, and P. Zoller, Phys. Rev. Lett. \textbf{105}, 015702 (2010).
\bibitem{Lee2013} T. E. Lee, S. Gopalakrishnan, and M. D. Lukin, Phys. Rev. Lett. \textbf{110}, 257204 (2013).
\bibitem{Jin2013} J. Jin, D. Rossini, R. Fazio, M. Leib, and M. J. Hartmann, Phys. Rev. Lett. \textbf{110}, 163605 (2013).
\bibitem{Weimer2015} H. Weimer, Phys. Rev. Lett. \textbf{114}, 040402 (2015).
\bibitem{Cui2015} J. Cui, J. I. Cirac, and M. C. Ba\~{n}uls, Phys. Rev. Lett. \textbf{114}, 220601 (2015).
\bibitem{Zanardi2015} P. Zanardi and L. C. Campos Venuti, Phys. Rev. A \textbf{91}, 052324 (2015).
\bibitem{Zhang2016} D.-J. Zhang, H.-L. Huang, and D. M. Tong, Phys. Rev. A \textbf{93}, 012117 (2016).
\bibitem{Lieb1974} E. H. Lieb and M. B. Ruskai, Adv. Math. \textbf{12}, 269 (1974).
\bibitem [{Note()}]{Note}%
  \BibitemOpen
  \bibinfo {note} {To see this, note that $\rho_0-\epsilon\rho$ is positive semidefinite for any density operator $\rho$ supported on $P_{\rho_0}\mathcal{H}$ and small enough number $\epsilon>0$. Employing the CP property of $\mathcal{E}_t$ and using the equality $\mathcal{E}_t(\rho_0)=\rho_0$, we have that $\rho_0-\epsilon\mathcal{E}_t(\rho)$ is also positive semidefinite. It is true only if $\mathcal{E}_t(\rho)$ is a density operator supported on $P_{\rho_0}\mathcal{H}$, thus implying the invariance of $P_{\rho_0}\mathcal{H}$ under $\mathcal{E}_t$.
  }\BibitemShut {Stop}%
\bibitem{Nielsen2000} M. A. Nielsen and I. L. Chuang, \textit{Quantum Computation and Quantum Information} (Cambridge University Press, Cambridge, England, 2000).
\bibitem{Lindblad1976} G. Lindblad, Commun. Math. Phys. \textbf{48}, 119 (1976); V. Gorini, A. Kossakowski, and E. C. G. Sudarshan, J. Math. Phys. (N.Y.) \textbf{17}, 821 (1976).
\bibitem{Davidson1996} K. Davidson, \textit{C*-algebras by example, fields institute monographs} (Amer. Math. Soc. Providence, 1996).
\bibitem{Zanardi2001} P. Zanardi, Phys. Rev. Lett. \textbf{87}, 077901 (2001).
\bibitem{Zanardi2004} P. Zanardi, D. A. Lidar, and S. Lloyd, Phys. Rev. Lett. \textbf{92}, 060402 (2004).
\bibitem{Baumartner} B. Baumgartner, H. Narnhofer, and W. Thirring, J. Phys. A: Math. Theor. \textbf{41}, 065201 (2008); B. Baumgartner and H. Narnhofer, \textit{ibid.} \textbf{41}, 395303 (2008).
\bibitem{Holbrook2003} J. A. Holbrook, D. W. Kribs, and R. Laflamme, Quantum. Inf. Proc. \textbf{2}, 381 (2003).
\bibitem{Chruscinski2010} D. Chru\'{s}ci\'{n}ski and A. Kossakowski, Phys. Rev. Lett. \textbf{104}, 070406 (2010).

\end{thebibliography}
\end{document}